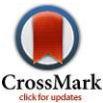

# Non-Scanning Fiber-Optic Near-Infrared Beam Led to Two-Photon Optogenetic Stimulation *In-Vivo*

Kamal R. Dhakal[1]⁹, Ling Gu[1]⁹, Shivaranjani Shivalingaiah[1], Torry S. Dennis[2], Samara A. Morris-Bobzean[2], Ting Li[3], Linda I. Perrotti[2], Samarendra K. Mohanty[1]*

1 Biophysics and Physiology Lab, Department of Physics, The University of Texas at Arlington, Arlington, Texas, United States of America, 2 Department of Psychology, The University of Texas at Arlington, Arlington, Texas, United States of America, 3 Key lab for Neuroinformatics of Ministry of Education, University of Electronic Science & Technology of China, Chengdu, Sichuan, China

## Abstract

Stimulation of specific neurons expressing opsins in a targeted region to manipulate brain function has proved to be a powerful tool in neuroscience. However, the use of visible light for optogenetic stimulation is invasive due to low penetration depth and tissue damage owing to larger absorption and scattering. Here, we report, for the first time, in-depth non-scanning fiber-optic two-photon optogenetic stimulation (FO-TPOS) of neurons *in-vivo* in transgenic mouse models. In order to optimize the deep-brain stimulation strategy, we characterized two-photon activation efficacy at different near-infrared laser parameters. The significantly-enhanced in-depth stimulation efficiency of FO-TPOS as compared to conventional single-photon beam was demonstrated both by experiments and Monte Carlo simulation. The non-scanning FO-TPOS technology will lead to better understanding of the *in-vivo* neural circuitry because this technology permits more precise and less invasive anatomical delivery of stimulation.

**Citation:** Dhakal KR, Gu L, Shivalingaiah S, Dennis TS, Morris-Bobzean SA, et al. (2014) Non-Scanning Fiber-Optic Near-Infrared Beam Led to Two-Photon Optogenetic Stimulation *In-Vivo*. PLoS ONE 9(11): e111488. doi:10.1371/journal.pone.0111488

**Editor:** Maurice J. Chacron, McGill University, Canada

**Received** June 10, 2014; **Accepted** September 24, 2014; **Published** November 10, 2014

**Copyright:** © 2014 Dhakal et al. This is an open-access article distributed under the terms of the Creative Commons Attribution License, which permits unrestricted use, distribution, and reproduction in any medium, provided the original author and source are credited.

**Data Availability:** The authors confirm that all data underlying the findings are fully available without restriction. All relevant data are within the paper and its Supporting Information files.

**Funding:** SKM is thankful for the support from Office of President and Provost, The University of Texas at Arlington, the National Institute of Health (NS084311), and Mr. Harvey Wiggins (Plexon Inc.) for providing the Plexon Omniplex system and Plexon staff for electrophysiological recording support. The funders had no role in study design, data collection and analysis, decision to publish, or preparation of the manuscript.

**Competing Interests:** The authors have the following interests: Mr. Harvey Wiggins (Plexon Inc.) provided the Plexon Omniplex system (demo unit) and Plexon staff for electrophysiological recording support. This does not alter the authors' adherence to PLOS ONE policies on sharing data and materials.

* Email: smohanty@uta.edu

⁹ These authors contributed equally to this work.

## Introduction

While neuronal imaging has enabled the mapping of the physical connectome [1] of the brain in a high-throughput manner, there is now a growing need for an efficient, high resolution method to functionally map the multitude of neural connections in the brain. However, until recently, such a task was challenging not only in terms of the non-invasiveness, but also in the required cellular specificity in the sea of multifunctional neural units. Recently, visible light-assisted activation of selected neuronal group has been made possible with high temporal precision by introducing a light-activated molecular channel; channelrhodopsin-2 (ChR2) [2,3,4,5] and its variants. This method has several advantages over electrical stimulation [6,7,8] such as cellular specificity and minimal invasiveness and therefore, growing as a valuable tool in neuroscience research. Though the use of light-activated molecular ChR2 for *in-vivo* models is increasing, limitations of its utility exist. Although optogenetic methods require very low intensity light (a few mW/mm²) [3], at the activation wavelength for ChR2 (460 nm) significant absorption and scattering of stimulating light occurs which leads to limited (shallow) penetration of the beam [9] (Fig. 1a). Thus, in order to stimulate cells in the most ventral regions of the brain, one has to choose between one of two undesirable alternatives: The first is to maintain the minimally invasive qualities of the approach but requires that the average beam power be significantly increased. Unfortunately, this approach using single photon (visible) light often leads to significant deleterious effects on the cell viability in the vicinity of the target, and thus may limit the translational potential of this technology. The second alternative approach is to use optical fiber [10,11,12] or μLED [13,14]-based blue light delivery to excite neurons in close proximity to the optical fiber or μLED. This approach may also compromise the minimally invasive qualities of optogenetics because optical fibers (similar to electrodes used in electrical stimulation techniques) or μLED(s) need to penetrate through more superficial brain tissue in order to reach more ventral brain regions. In this way, the more superficial brain tissue gets damaged; and such damage can lead to difficulty in interpreting the outcome of deep-brain stimulation. Therefore, several attempts have been made to shift the activation peak of ChR2 from the blue to the red wavelength region [15,16,17]. While there has been some success at the cost of altered light-activation kinetics, it seems a near-infrared opsin would be ideal for in-depth cell-specific optogenetic stimulation of excitable cells in an organism. However, the success in developing a near-infrared opsin would still not permit applicability of the method for selective activation of small regions in deep (ventral) brain areas, since the single-photon (red or infrared) stimulating beam will also stimulate neurons outside of the targeted population (in the path of the beam, Fig. 1b).

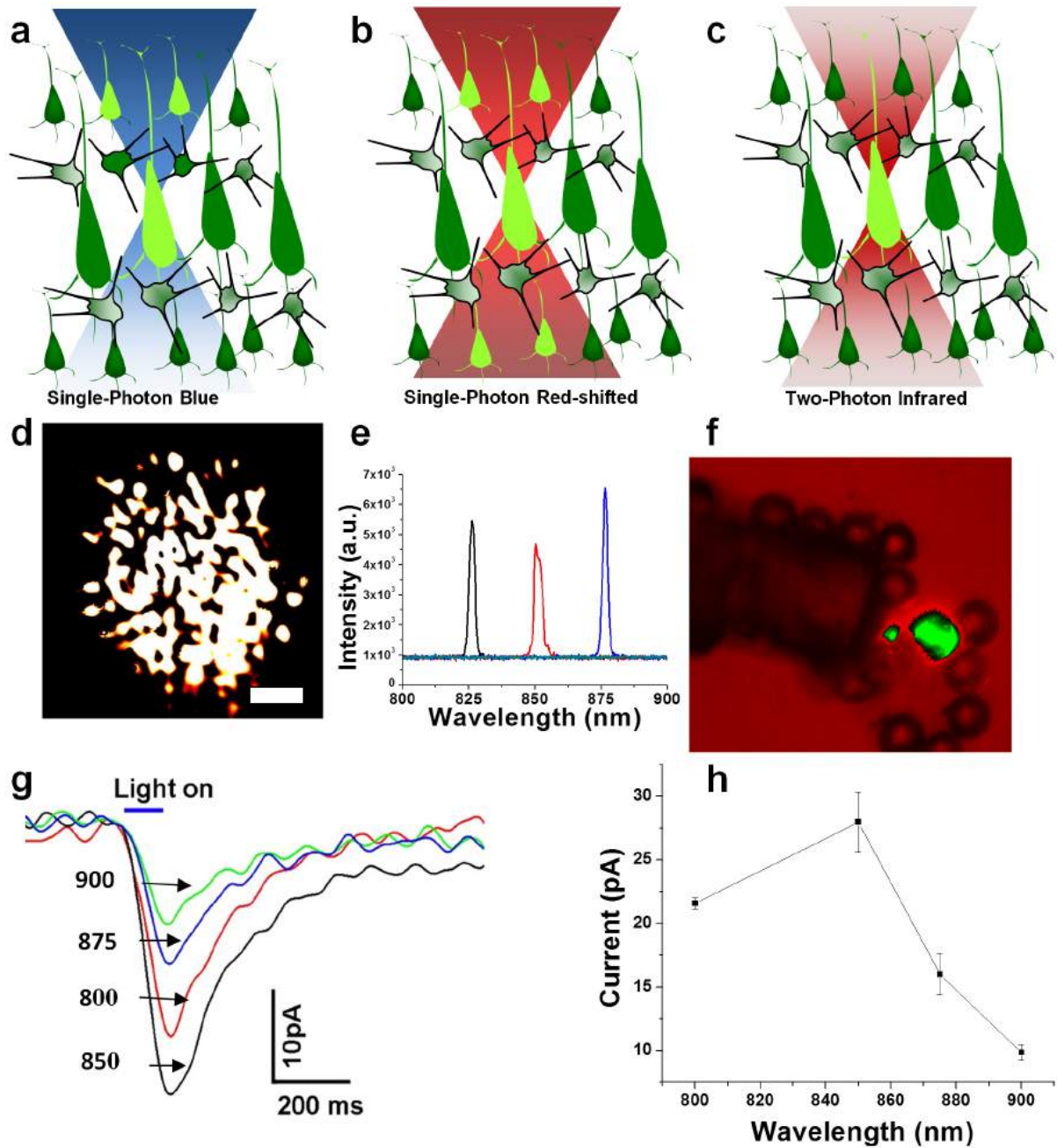

**Figure 1. Two-photon Optogenetic stimulation.** Schematic comparison: (a) single-photon blue excitation, (b) single-photon red-shifted excitation and (c) two-photon optogenetic stimulation (TPOS). (d) Intensity profile emanating from the multimode fiber, scale bar: 100 µm, (e) Spectrum of ultrafast tunable near-infrared fiber-optic beam, (f) Two-photon fluorescence (green) from polystyrene particle excited by multimode profile. In-vitro two-photon activation of ChR2-expressing cells with ultrafast NIR laser beam: (g) Representative current responses to ultrafast NIR laser beam at different wavelengths (in nm), (h) Fiber-optic two-photon activation spectrum at 0.02 mW/µm² (100 ms pulses).
doi:10.1371/journal.pone.0111488.g001

Because of low absorption and scattering coefficients, the two-photon optogenetic activation using a near infrared laser beam can provide both deep penetration and high spatial precision, achieved by virtue of the non-linear nature of ultrafast light interaction with ChR2. Comparison of single-photon blue (a), and red-shifted excitation (b) with two-photon optogenetic stimulation (c) is shown in Fig. 1. Since the first demonstration [18] of *in-vitro* two-photon optogenetic stimulation (TPOS) of ChR2-sensitized cells and neurons by using point or scanning laser beam, there has been significant progress [19,20,21,22] in using TPOS for probing neural circuitry in cultured neuronal cells as well as brain slices. Further, recent advances demonstrate that optogenetically-sensitized neurons can be activated by spatially-sculpting [20] and/or temporal-focusing [21] the TPOS beam. It is important to note that the two-photon excitation has significant benefits over the single-photon excitation such as less photo-damage, enhanced penetration depth and localized stimulation [23]. Further, two-photon cross-section of ChR2 is empirically estimated [19]' [21] to be larger than that of most of the fluorophores and therefore, has potential for stimulating opsin-expressing neurons using weakly-focused (and even defocused) two-photon beams. However, to date, two-photon excitation has only been demonstrated *in-vitro* with microscope-based setups [18,19,24] and recently by fiber-optic stimulation [25]. Here, we demonstrate *in-vivo* fiber-optic two-photon optogenetic stimulation (FO-TPOS) of in-depth neuronal circuitry in transgenic mouse model. Our experiments, reported here, demonstrate that multimode fiber can be used: (1) *in-vivo* as a non-scanning beam unlike that used in earlier microscope objective based two-photon stimulation reports, (2) to achieve in-depth two-photon activation *in-vivo* and (3) to evaluate intensity and wavelength-dependent *in-vivo* fiber-optic two-photon optogenetic stimulation efficacy. Here, we would like to emphasize that this manuscript does not examine the axial localization ability of two-photon stimulation *in-vivo*. Stimulation of ensemble of chemically-identical cells at large depths by two-photon fiber-optic optogenetic means as demonstrated here, in a non-invasive manner will enable dissecting the role of those ensembles in particular function and thus modulate behavior.

## Materials and Methods

### Ethics statement

All experimental procedures were conducted according to the University of Texas at Arlington-Institutional Animal Care and Use Committee approved protocol A10.009. The IACUC specifically approved this study.

### Cell culture

HEK 293 cells were transfected with the ChR2-EYFP construct, cloned into pcDNA3.1 neo (Invitrogen, USA). ChR2 cDNA was kindly provided by Dr. Georg Nagel (University of Wuerzburg, Germany). EYFP was fused in-frame to the C-terminus of ChR2 by PCR. Transgene-expressing cells were identified by visualizing the EYFP fluorescence under suitable illumination (514 nm). Stable clones were selected with 200 mg/l G418 and colonies were picked after 2 weeks, and then expanded. Clones that showed the highest level of EYFP fluorescence were chosen for the optical activation experiments. Cells were maintained at $37^{\circ}C$, 5% $CO_2$ in DMEM containing 10% fetal bovine serum. For generating light activation, cells were loaded with all-trans retinal (1 $\mu M$) for at least 6 hours and activated with fiber-optic laser beam.

### Optogenetic stimulation

For both the *in-vitro* and *in-vivo* activation of ChR2-expressing cells (identified by YFP fluorescence) or different brain regions expressing ChR2, both two-photon and single photon sources coupled to a 50 $\mu m$ core optical fiber using a fiber coupler (Newport Inc.), mounted on a mechanical micromanipulator so as to position the tip of the fiber near the region of interest. The single-photon source consists of a blue (473 nm, 30 mW) diode laser coupled to the fiber, while the two-photon beam is provided by a 200 fs near-infrared Ti: Sapphire laser (Maitai HP, Newport-SpectraPhysics Inc.) beam operating at ~76 MHz, coupled to the same fiber. Further, the two-photon laser wavelength was tuned from 750 nm to 950 nm in order to compare the relative efficiencies of the various NIR wavelengths. Macro-exposure pulses of stimulation light were generated by a function generator driving the electro-mechanical shutter in the beam path. The function generator was synchronized to the *in-vitro* and *in-vivo* electro-physiology recording system (Plexon and Biopac Inc.). Light power at the fiber-tip was measured using a standard light power meter (PM 100D, Thorlabs Inc).

### Patch-clamp recording setup

The opto-electrophysiology set up was developed on an Olympus upright microscope platform using an amplifier system (Axon Multiclamp 700B, Molecular Devices Inc.). Parameters of the pipette puller were optimized in order to obtain desired borosilicate micropipettes of resistance from 3 to 5 $M\Omega$ for whole-cell patch clamp. The micropipette was filled with a solution containing (in mM) 130 K-Gluoconate, 7 KCl, 2 NaCl, 1 $MgCl_2$, 0.4 EGTA, 10 HEPES, 2 ATP-Mg, 0.3 GTP-Tris and 20 sucrose. The electrode was mounted on a XYZ motorized micromanipulator (Newport Inc.). The standard extracellular solution containing (in mM): 150 NaCl, 10 Glucose, 5 KCl, 2 $CaCl_2$, 1 $MgCl_2$ was buffered with 10 mM HEPES (pH 7.3). The output from the micropipette was digitized using a National Instruments card (PCI 6221). For electrophysiological recording, the hardware was interfaced with patch-clamp software from University of Strathclyde (non-commercial use). Electrical recordings were performed at a holding potential of $-60$ mV at room temperature ($20-24^{\circ}C$). For activation of ChR2-expressing cells (identified by YFP fluorescence), the optogenetic stimulation beam (473 nm, or NIR) was delivered via a 50 $\mu m$ core optical fiber, mounted on a mechanical micromanipulator so as to position the tip of the fiber near the desired cell being patch-clamped. For generating and controlling pulses of light, the electromechanical shutter in the laser beam path was interfaced with a PC. TTL pulses of desired frequency were generated using National Instruments (PCI 6221) card in order to generate required laser pulses for activation. For electrophysiological measurements subsequent to optical activation, the shutter was synchronized with the patch clamp recording electrode. The whole system was built on a vibration isolation table (Newport Inc.) and electrical isolation was done by means of a Faraday cage that was placed around the setup. pClamp software was used for data analysis.

### Monte Carlo Simulation

Monte Carlo simulation is known as the most reliable and flexible method for modeling photon migration in biological tissue. Here, we used the widely-used Monte Carlo simulation codes for light propagation in multi-layered biological tissue named 'MCML' [26] and 'COV' [27]. The brain was modeled as a two-layered tissue including grey matter and white matter. We set up parameters [28,29,30] ($\mu_a$ = absorption coefficient; $\mu_s$ = scattering coefficient; g = anisotropy factor; n = refractive index, and d = thickness of cortical layers) as listed in Table S1 for the simulation, and then computed light propagation with the MCML code. Then the paired code 'COV' was utilized to extract the simulation output and calibrate the computed fluence rate distribution by quantitatively carrying the effect of Gaussian beam distribution and respective parameters, including beam size, laser power, and numerical aperture (NA). $10^7$ photons were launched in each simulation to achieve an excellent signal-to-noise ratio in simulation output [26,27].

### Mouse preparation

The (Thy1-ChR2-YFP) transgenic (n = 6, age between 12 to 16 weeks), and wild-type (C57BL/6J) mice (n = 3, age between 12 to 16 weeks) used in the reported experiments, were purchased from Jackson Labs. All aspects of experimental manipulation of our animals were in strict accordance with guidelines of the University of Texas at Arlington's Institutional Animal Care and Use Committee (IACUC). Mice were maintained on a 12:12 light cycle (lights on at 07:00).

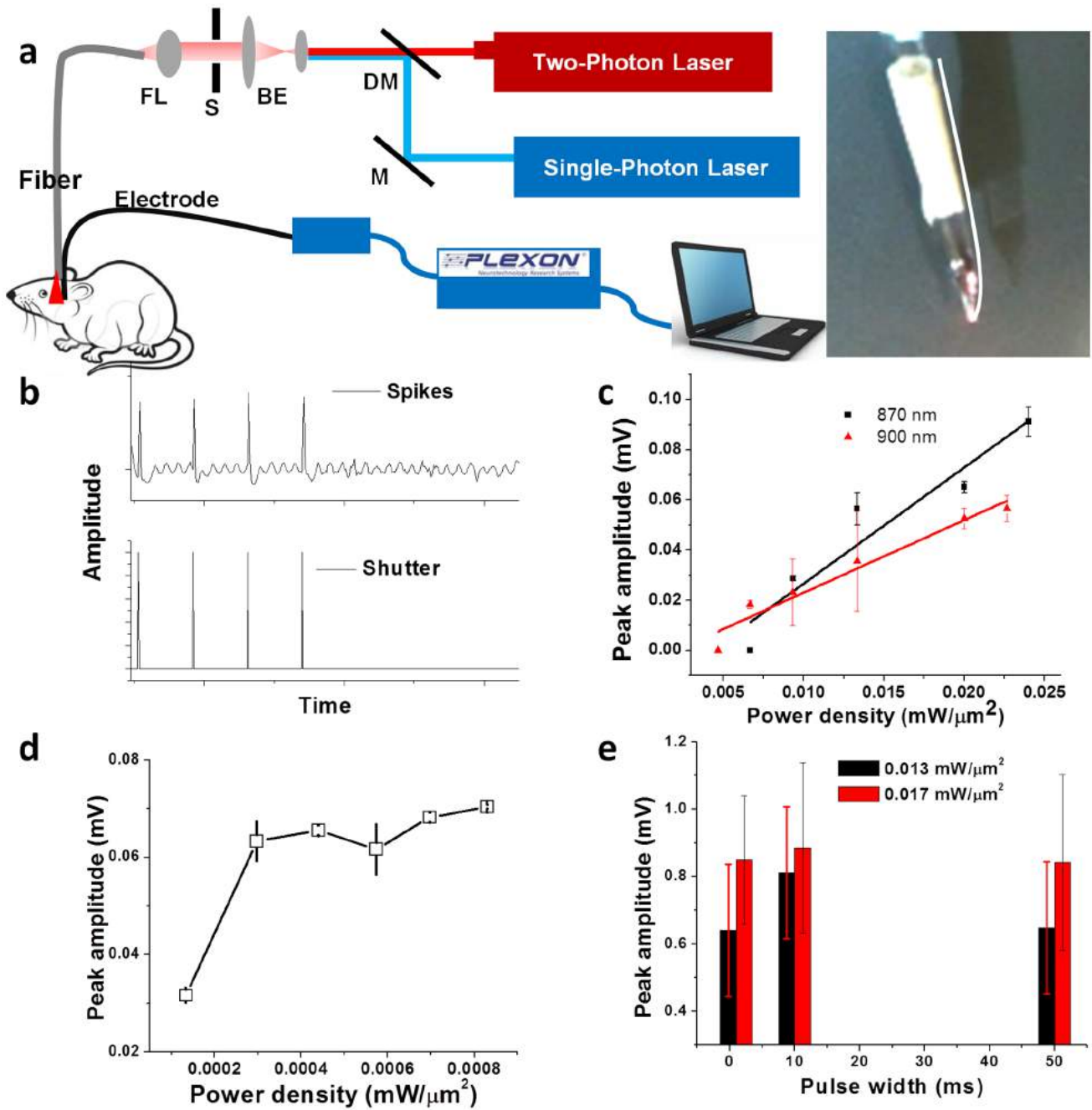

**Figure 2. In-vivo fiber-optic optogenetic stimulation using single (blue) vs two-photon (NIR) stimulation.** (a) Set-up for in-vivo single and two-photon fiber-optic optogenetic stimulation and electrophysiological recording. Inset in right: Glass electrode and fiber based optrode. (b) In-vivo raw spiking of locally stimulated neurons (upper panel) corresponding to two-photon stimulation (shutter opening: lower panel) at 870 nm. (c) Peak-amplitude of recorded local electrical activity as a function of incident power density of in-vivo fiber-optic TPOS at two different wavelengths (also shown fitted lines). (d) Variation of peak amplitude of single photon (473 nm)-activated local electrical activity as a function of power density. (e) Peak-amplitude as a function of pulse width of in-vivo fiber-optic TPOS (870 nm) at two different power densities.
doi:10.1371/journal.pone.0111488.g002

### In-vivo electrophysiology recordings

Animals were deeply anesthetized with 90 mg/kg ketamine and 10 mg/kg xylazine and placed in a stereotaxic frame (Kopf Instruments Inc.). For in-vivo recording, a linear midline skin incision was made and burr holes were bilaterally drilled in the skull at the anteroposterior (AP) (in reference to bregma) and mediolateral (ML) coordinates corresponding to VTA (−3.2 mm AP; ±0.5 mm ML). The electrode was implanted stereotaxically to allow recording from deep brain regions. Continuous electrical activity was then recorded either via a multichannel acquisition system (Omniplex, Plexon Inc.) or via an amplifier (MP-150, Biopac Inc.) interfaced with the Acqknowledge acquisition

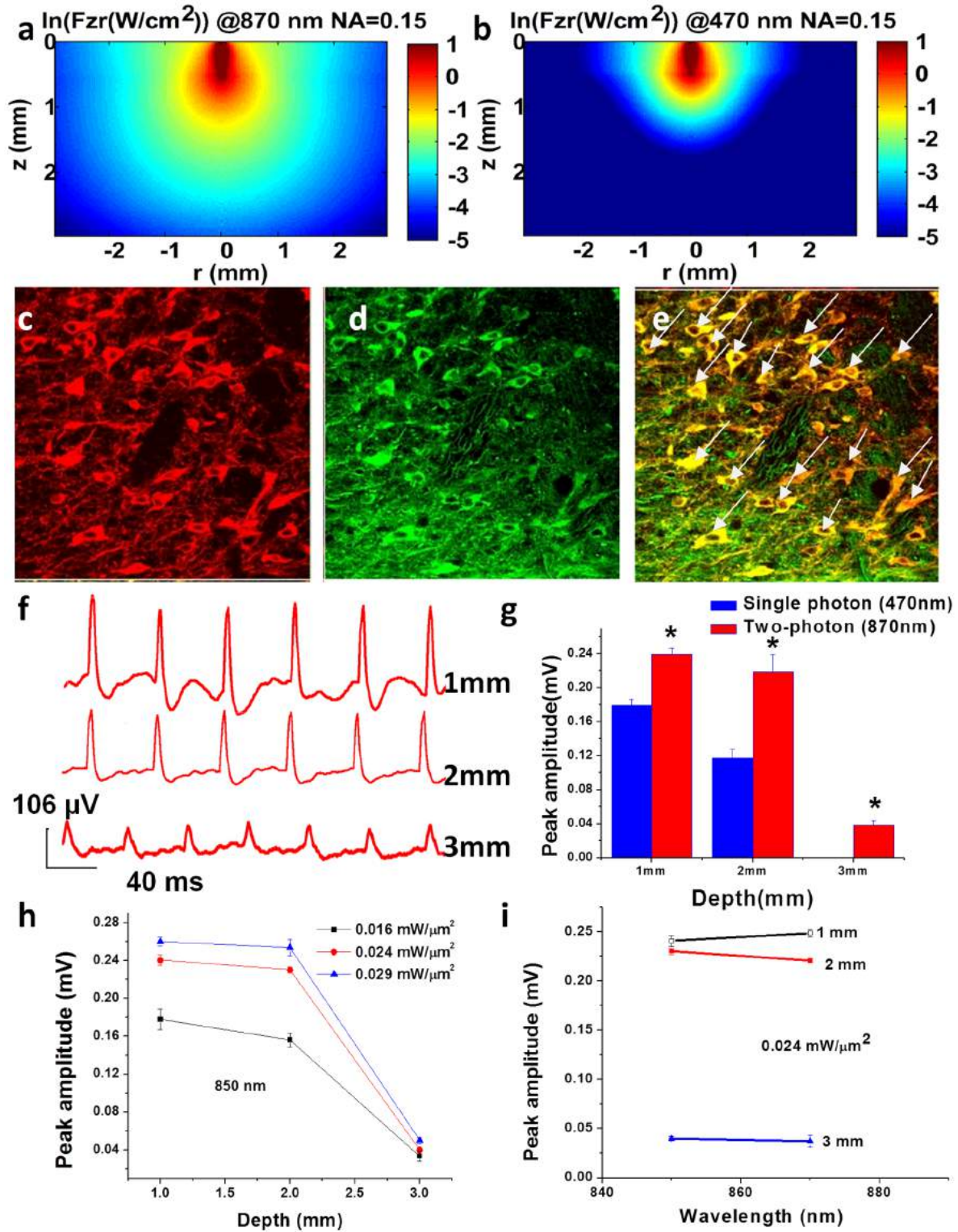

**Figure 3. In-depth fiber-optic optogenetic stimulation using single (blue) vs two-photon (NIR).** Monte Carlo simulation of light propagation in two-layered cortex for beam diameter of 60 μm, with laser power of 5 mW, (a) two-photon and (b) single-photon Gaussian beam. The color bar has unit of ln(W/cm$^2$). Confocal immunofluorescence images of a deep brain region of Thy1-ChR2-YFP transgenic mice: (c) YFP, (d) tyrosine hydroxylase (TH), (e) composite image. Arrows point to examples of colocalized YFP-ChR2 in dopaminergic neurons. (f) Raw spiking activity at different depths due to *in-vivo* fiber-optic two-photon optogenetic stimulation (FO-TPOS). (g) Comparison of single and two-photon activated depth-dependent peak amplitude. *: p<0.05 vs. single photon. (h) Variation of peak amplitude of two-photon (850 nm)-activated local electrical activity as a function of depth at different power densities. (i) Comparison of peak-amplitude at two different wavelengths of *in-vivo* fiber-optic TPOS at different depths.

doi:10.1371/journal.pone.0111488.g003

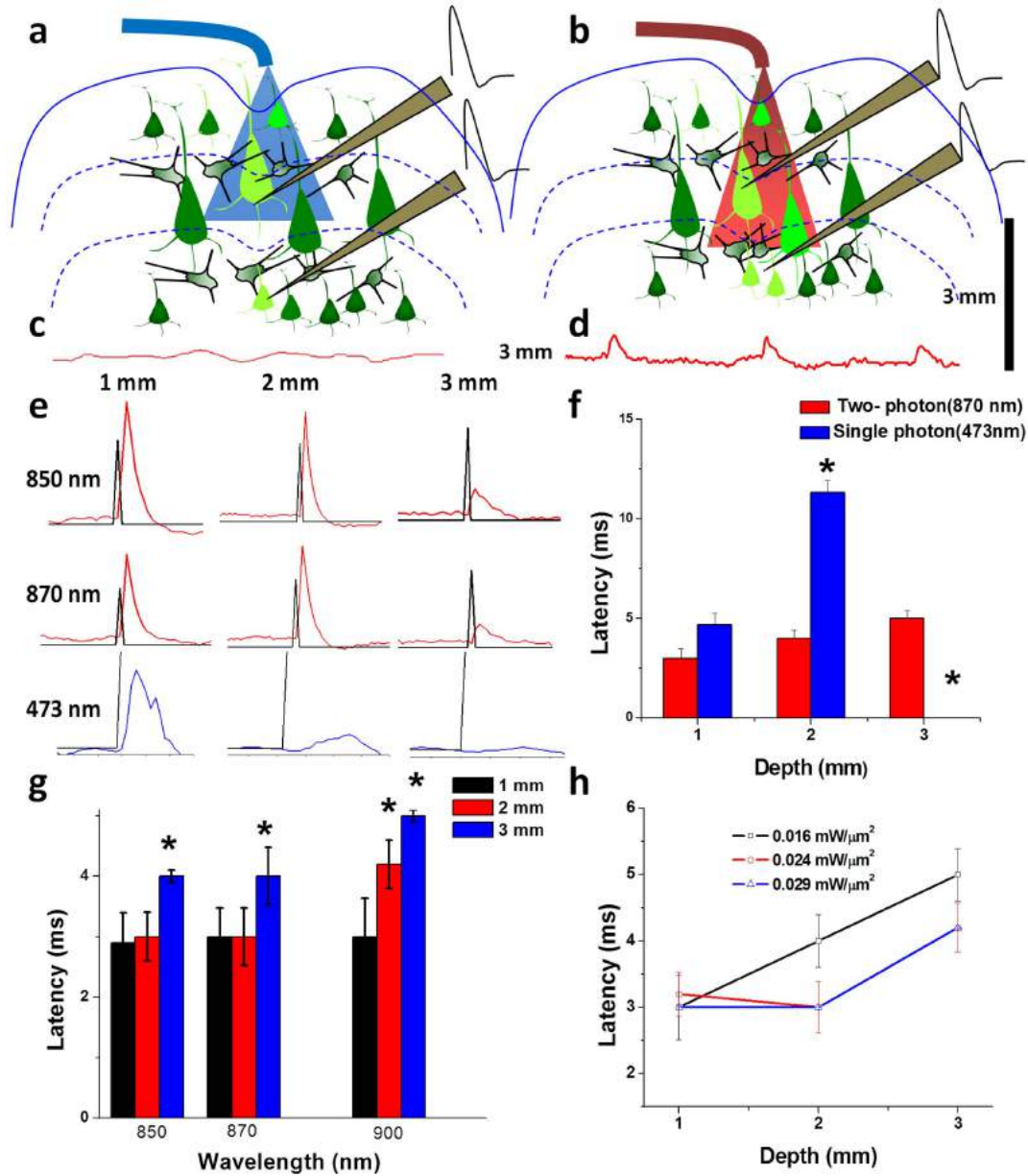

**Figure 4. Depth-dependent *in-vivo* fiber-optic single and two-photon optogenetic stimulation.** Sketch showing in-depth stimulation limit of fiber-optic (a) blue (single-photon) vs (b) NIR (two-photon) light. Different cortical layers are marked by blue lines. Raw electrical recording at 3 mm depth for (c) single photon (473 nm, 4 mW) and (d) two-photon (870 nm, 80 mW) stimulation. (e) Overlay of raw spikes (red profiles) with shutter (fiber-optic two-photon stimulation ON) opening (black profiles) at different depths for two different wavelengths. Also shown are representative single photon signals (blue profiles) at different depths, with shutter-ON marked by black profiles. (f) Comparison of latency of spikes at different depths stimulated by single photon and two-photon (870 nm) fiber-optic beams. *: p<0.05. (g) Histogram of latency of spikes at different two-photon wavelengths for three depths. *: p<0.05 vs. latency at 1 mm. (h) Variation of latency as a function of depth for different two-photon (870 nm) power densities.
doi:10.1371/journal.pone.0111488.g004

software on a networked PC. The animals were euthanized following electrophysiology recoding. by intraperitoneal injection of Sodium Pentobarbital (100 mg/kg).

## Statistics

Data were analyzed using Statistical Package for the Social Sciences (SPSS, IBM). One-way analysis of variance (ANOVA) followed by Tukey's post-hoc test was used to determine whether a significant difference occurred. The data were plotted as mean ± S. D. The accepted level of significance was $p<0.05$.

## Results

### *In-vitro* fiber-optic near-infrared ultrafast laser irradiation leads to optogenetic stimulation

In contrast to the use of a microscope objective for focusing the two-photon laser beam, we employed a tunable (800–900 nm), ultrafast (200 fs) fiber-optic beam for non-scanning two-photon optogenetic activation. It is important to note here that earlier TPOS methods utilized different scanning modes (spiral, raster) for activation of ChR2-expressing cells to optimize the efficiency of excitation [19]. For testing the efficiency of fiber-based non-scanning two-photon activation *in-vitro*, HEK-293 cells expressing ChR2 were used and channel activities were recorded with patch clamp. The patch clamp set up for electrophysiological measurement subsequent to FO-TPOS is shown in Fig. S1a. The tunable Ti: Sapphire laser (FSL, 76 MHz) beam was expanded using a beam expander (BE) and coupled to an optical fiber using a lens (L). A shutter (S) controlled the exposure (macro-pulse) duration and a circular neutral density filter (NDF) was used to control the average beam power. In the case of single mode fiber-optic beam or laser beam focused by microscope objective, the intensity and, therefore, the two-photon stimulation is maximum at the center and slowly decays towards periphery (Gaussian nature). However, in case of the multimode fiber-optic beam, as used here, the maximum intensity spots are distributed over the irradiated cell(s) and two-photon stimulation is expected to occur in these spots leading to non-scanning activation of the cell(s). Fig. 1d shows the intensity profile of the NIR (875 nm) beam emanating from the multimode fiber at a distance of ~1.3 mm from the cleaved end. In Fig. S1(e), we show the time-lapse (1 sec) intensity profiles (in different colors) along a line drawn across the beam profile (Fig. S1d). The beam profile was found to be stable (average correlation coefficient between two frames is 0.98) unless the fiber was mechanically actuated externally (Movie S1). In order to estimate the pulse width of the ultrafast NIR fiber-optic beam, the spectrum was measured using spectrometer (Ocean Optics Inc) at different wavelengths (Fig. 1e). At a fixed wavelength, variation of intensity (by changing the pump laser intensity) led to change in pulse width at lower pump power as shown in Fig. S1b. Since no significant change in pulse width at high oscillator power (1.3 to 1.7 W) was observed, this range was used for intensity-dependent studies. This range of oscillator IR laser power corresponds to ~150 to 260 mW at the coupling end of the multimode fiber. With a coupling efficiency of nearly 50%, the average power at tip of the fiber was measured to be 75 to 130 mW. Before evaluating two-photon activation of cells by multimode fiber-optic beam, we examined if the ultrafast fiber-optic near-infrared beam has sufficient photon densities to result in two-photon excitation. Therefore, we carried out two-photon excitation of fluorescent polystyrene particles (Bangs Lab) by placing the cleaved fiber tip near the particles. The beam spot size is about 50 μm when the fiber was placed 10 μm away from the particles. Fig. 1f shows the green fluorescence emitted from 45 μm polystyrene particle(s) excited by the multimode fiber-optic beam (Movie S2). Non-linear nature of the incident intensity-dependent fluorescence emission confirmed role of two-photon excitation. It may be noted that as distance of the fluorescent particle from the fiber increases, the excitation intensity decreases due to divergence of the beam from the fiber and therefore, leads to change in excitation volume.

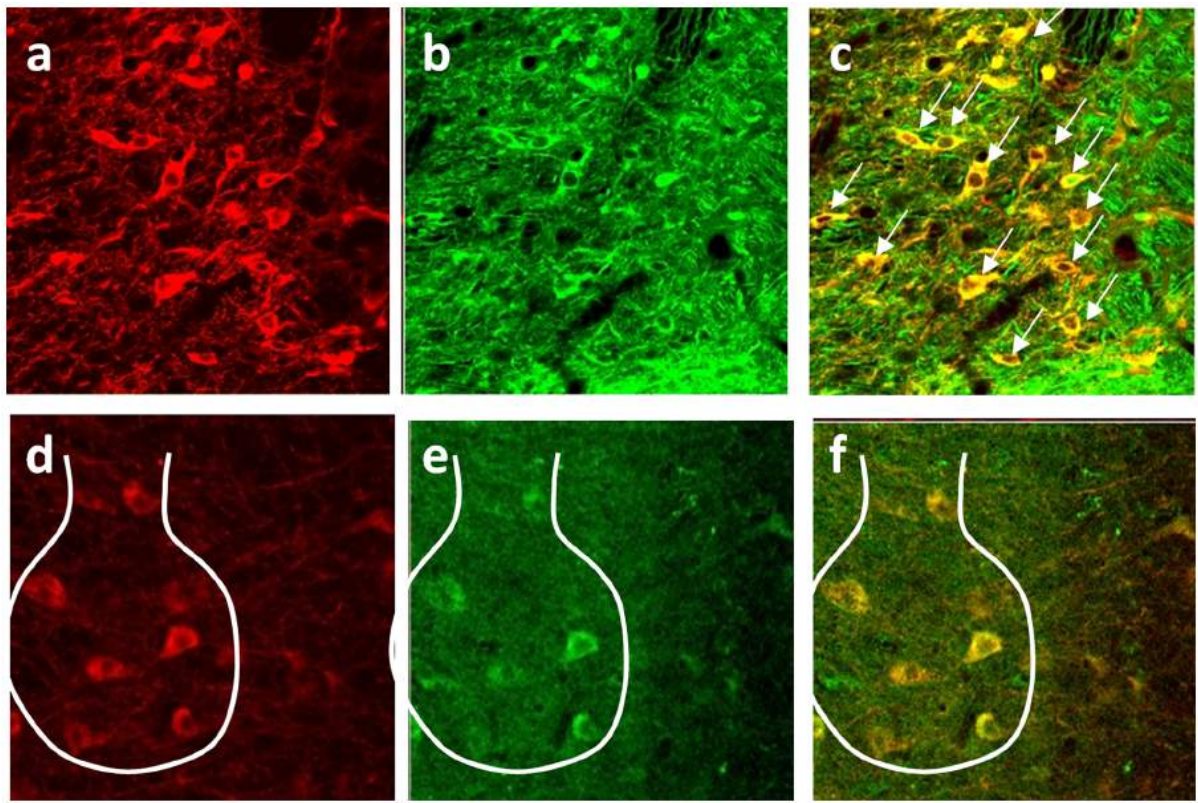

**Figure 5. In-depth c-Fos expression in dopinergic cells due to fiber-optic two-photon optogenetic stimulation.** Confocal immunofluorescence images of ventral tegmental area (VTA) of Thy1-ChR2-YFP transgenic mice. (a) YFP, (b) tyrosine *hydroxylase* (TH), and (c) Composite image showing co-localization of YFP and TH. Arrows point to examples of co-localized cells. Increase in c-Fos expression of dopinergic cells in VTA due to fiber-optic two-photon optogenetic stimulation (FO-TPOS); (d) TH, (e) c-Fos, and (f) Composite image showing co-localization of TH and c-Fos. Few cells have been marked as predicted (MC simulation) by the boundary of the FO-TPOS beam profile.
doi:10.1371/journal.pone.0111488.g005

In order to map the *in-vitro* fiber-optic two-photon activation spectrum of ChR2, the wavelength of the near-infrared stimulation laser beam was tuned from 800 to 900 nm. Expression of ChR2 in HEK 293 cells was confirmed by fluorescence imaging of reporter fluorescent protein (YFP). The cleaved multimode fiber tip was positioned ~100 μm away from the clamped-cell. In Fig. 1 g, we show representative current responses recorded from a single cell, irradiated with ultrafast NIR laser beam (100 ms pulses, with average power density of 0.02 mW/μm², at cell membrane) at different wavelengths. The fiber-optic two-photon activation spectrum of ChR2 (Fig. 1 h) *in-vitro* shows that the peak activation wavelength is at 850 nm. Theoretically, two-photon optogenetic excitation is expected to be maximal near 920 nm (double the single photon excitation peak wavelength). The current amplitude at 900 nm was 9.9±0.6 pA (n = 18), which was lower than the current amplitude at 850 nm (28.0±2.4 pA, n = 19). The observed blue-shift of the two-photon activation spectrum can be attributed to the parity selection rules favoring the two-photon excitation of all-trans-retinal and thus activation of ChR2 (in closed form) to higher energy (open) state than the respective single-photon-induced activation. This "blue-shifted" optimum stimulation wavelength for two-photon excitation with respect to single-photon excitation is similar to that observed [31,32] in two-photon absorption of many fluorescence dyes and in two-photon activation ChR2-sensitized HEK cells [20].

## Estimation of two-photon absorption cross section

For estimation of fiber-optic two-photon cross-section of ChR2-activation (Fig. S2), we have considered the dispersion-induced measured pulse-width. We estimated the two-photon absorption cross section by fitting the equation to the inward current obtained from patch-clamp as shown in Fig. S2 (a–b). The inward increasing current was fitted with following equation,

$$
I^*(T) = I^*(\max)\{\frac{\tau_2}{\tau_g - \tau_2}(e^{-\frac{T}{\tau_g}} - e^{-\frac{T}{\tau_2}}) \\
- \frac{\tau_1}{\tau_g - \tau_1}(e^{-\frac{T}{\tau_g}} - e^{-\frac{T}{\tau_1}})\}
\tag{1}
$$

Where $\tau_g$ is the ground state life time, defined $(2/\sigma_2 \, I^2)$ for two photon excitation and $(1/\sigma_1 \, I)$ for single photon excitation. $I$ = intensity, $\sigma_2$ = two-photon absorption cross-section, $\tau_1$ = activation time-constant, and $\tau_2$ = current decay time-constant.

From the fitted parameters the two photon absorption cross-section at $\lambda = 870$ nm and 900 nm was found to be 226.48±5.034 GM and 186.32±33.03 GM respectively. This estimated two-photon absorption cross-section is higher than many known fluorophores and is comparable to earlier report [19]. This is one of the reasons why fiber optics could produce two-photon activation of ChR2 at low average intensities.

## Peak of *in-vivo* fiber-optic two-photon stimulation wavelength is also blue-shifted

In order to achieve *in-vivo* two-photon stimulation of different brain regions of transgenic mice (Thy1-ChR2-YFP), fiber-optic delivery (Fig. S1c) of the fs laser beam was employed instead of microscope objective. The fiber tip was placed in the cortical layer-1 of the transgenic mice and the electrode was translated stereotactically to 1 mm beneath the fiber tip. In case of fiber-optic two-photon stimulation, the laser beam irradiates the entire cell(s). In order to compare the relative efficiencies of various NIR wavelengths in fiber-optic two-photon optogenetic stimulation (FO-TPOS) *in-vivo*, the (Ti: Sapphire) laser wavelength was tuned from 800 to 900 nm, maintaining same power levels. Fig. S3a shows spiking activity recorded in cortical regions by *in-vivo* electrophysiology subsequent to fiber-optic two-photon optogenetic stimulation at three selected wavelengths (average incident laser power density: 0.013 mW/μm², pulse width: 50 ms; frequency of shutter: 5 Hz). Fig. S3b shows variation of peak voltage as a function of *in-vivo* FO-TPOS wavelength. Though the *in-vivo* two-photon activation spectrum is of similar nature to that of the in-vitro patch-clamp experiments (Fig. 1), the slope was less steep towards higher activation wavelengths. At these NIR three wavelengths, the average power density of the fiber-optic beam was increased to 0.023 mW/μm², and spikes were recorded from the same site (n = 5). Fig. S3b shows an increase in peak amplitude of the recorded spikes at all the three wavelengths of *in-vivo* FO-TPOS for higher laser power density (p<0.02, n = 5).

In order to further evaluate the efficacy of wavelength-dependent fiber-optic *in-vivo* two-photon activation, the firing rate (spikes per second) was measured and normalized (after subtracting number of background spikes, if any). Fig. S3c shows the normalized firing rate as a function of wavelength of *in-vivo* FO-TPOS. In the case of two-photon firing-rate based activation spectrum, the wavelength-dependency is more favorable towards longer wavelengths as compared to the peak-amplitude (Fig. S3b) and current based activation spectrum (Fig. 1 h). This may be attributed to differential absorption and scattering properties of neural tissue and water in the near-infrared region. While at longer wavelengths, the single-photon water absorption increases, the two-photon absorption of cellular chromophores and ChR2 is higher towards lower wavelengths. The fact that at higher wavelength (1000 nm), no significant stimulation was observed led us to conclude that contribution of direct (photothermal or photomechanical) stimulation is minimal. Further, these direct effects have earlier been found to occur at significantly higher pulse energies [33] and power densities [34] than that used in our current experiments.

## *In-vivo* fiber-optic two-photon optogenetic stimulation is intensity dependent

To determine the variation of *in-vivo* two-photon activation efficacy as a function of average laser power density (intensity), the ChR2-expressing cortical regions of the transgenic mice were stimulated using the setup shown in Fig. 2a. Further, to compare the intensity-dependency of *in-vivo* two-photon stimulation with that of single-photon, a blue diode laser (473 nm) was combined with the two-photon laser beam using a dichroic mirror (DM), co-aligned by a folding mirror (M), expanded by beam expander (BE) and coupled to a single fiber using a lens (FL). A computer-controlled shutter (S) was used (Fig. 2a) to control the stimulation pulse duration and the frequency of both the single and two-photon activation beams. In order to rule out photoactivation of the recording electrodes, which would lead to spurious signals, the optical fiber was affixed (shown by white line in right inset in Fig. 2a) to a glass micro-pipette with the fiber-tip positioned ~1 mm ahead of the pipette tip. A tungsten electrode was inserted into the pipette containing extracellular recording solution ~2 cm away from the fiber tip. Loss of some near-infrared light from the fiber can be seen near the shank of the pipette tip due to bending of the fiber (inset in Fig. 2a). Fig. S4a & b show recordings from the micropipette-electrode during ON and OFF cycles of ultrafast near-infrared irradiation in absence and presence of phosphate buffer saline respectively. The recording using the micropipette-electrode during *in-vivo* fiber-optic two-photon illumination of cortical region of deceased transgenic mouse brain is shown in Fig. S4c. The stimulation pulses (shutter opening and closing) are shown below all recordings in Fig. S4. Thus, the use of micropipette-electrode in combination with the light-delivery fiber being at fixed distance (1 mm) from the tip allowed us to remove the artifact (if any). In Fig. 2b, we show *in-vivo* raw spiking of locally stimulated ChR2-expressing cortical neurons (upper panel) of transgenic mouse in synchronization with two-photon stimulation (shutter opening: lower panel) at 870 nm. FO-TPOS experiments performed in ChR2 negative mice (n = 3, C57BL/6J) did not lead to any light-evoked electrical activity at depth of 1 mm for incident average intensity up to $0.02$ mW/$\mu$m$^2$ (870 nm, 80 MHz, shutter pulse width: 100 ms).

The micropipette-electrode-fiber carrying both blue and near-infrared laser beam was used to conduct an intensity-dependent comparison between single and two-photon in-vivo stimulation. Average incident laser power density was varied from ~0.006 to 0.026 mW/$\mu$m$^2$ for two-photon stimulation and that of single-photon stimulation was varied from ~0.0001 to 0.0008 mW/$\mu$m$^2$. The selection of power levels for comparison of two-photon with single-photon stimulation is based on previous in-vitro patch clamp measurements [20] which shows that two-photon (820 nm) intensity of ~100 times higher than that of single-photon (488 nm) is required to elicit similar photocurrent (~500 pA) in ChR2+ve HEK cells. Though the selected power-density range for two-photon activation is higher (~40 times) than that of single-photon, the phototoxicity data in cells shows that threshold average power density of near-infrared fs laser (810 nm) is ~1000 times higher than that due to cw blue (458 nm) laser beam [35]. Two-photon activation at such low intensities is believed to be possible due to the large two-photon cross section of ChR2 [19]. The highest average power density value (~2600 W/cm$^2$) for mode-locked near-infrared fiber-optic laser beam, used in the experiments reported here, is comparable to the reported [35] damage-threshold value of 1900 W/cm$^2$ for more-absorbing pigmented cells (with 250 ms exposure time). To further minimize damage in our intensity-dependent fiber-optic two-photon activation experiments, the pulse width of the shutter was set at 1 ms, thus resulting in a much lower dose than the damage threshold. Fig. 2c shows the peak-amplitude variation of recorded local electrical activity as a function of incident intensity of in-vivo FO-TPOS at two different wavelengths (870 nm and 900 nm) for same stimulation-detection sites (n = 4). The peak amplitude can be seen to increase with incident intensity for both wavelengths, but with a larger slope for 870 nm than 900 nm. This intensity-dependent trend is not quadratic and the peak amplitudes are significantly different only at power densities $\geq 0.02$ mW/$\mu$m$^2$. Besides the fact that conformational change of ChR2 upon two-photon irradiation is unlike excitation of fluorescent molecules, other parameters such as the kinetics of the opening of the ChR2-channel and two-photon absorption of all-trans-retinal [36,37] will significantly modulate the nature of intensity-dependent current variations. Further, the in-vivo non-quadratic intensity-dependent behavior can also be attributed to higher number of neurons being recruited during increase in intensity. Therefore, the measured intensity-dependent peak-amplitude in case of in-vivo FO-TPOS is not perfectly linear as observed in the intensity range studied for in-vitro [25].

The peak amplitude of single photon (473 nm)-activated neuronal firing as a function of power density is shown in Fig. 2d. In contrast to in-vivo two-photon stimulation, the single-photon stimulation led to a saturation of peak amplitude at a threshold average power density of $3 \times 10^{-4}$ mW/$\mu$m$^2$. This may be attributed to the fact that unlike two-photon activation, most of the units near the vicinity of the recording electrode, contributing to the signal amplitude, are already stimulated within the excitable volume (limited depth) of the single photon beam. The red-shifted, one-photon stimulation is also expected to cause similar saturation behavior as a single-photon blue laser beam. The larger dynamic-range (Fig. 2c) observed in the case of two-photon stimulation is advantageous in controlling the stimulation volume. This is owing to the non-linear nature of the two-photon activation process.

When the stimulation power density increased from 0.0065 to 0.013 mW/$\mu$m$^2$, the firing rate of neurons also increased significantly (p<0.01, n = 6). In order to further ascertain that two-photon stimulation using such small pulse-width (1 ms) is sufficient to generate spiking, the peak amplitude values were compared at three different pulse widths (1, 10, 50 ms). Fig. 2e shows the variation of peak-amplitude as a function of pulse width for in-vivo FO-TPOS (870 nm) at two different average power densities (0.013 and 0.017 mW/$\mu$m$^2$). While larger average power density led to higher peak amplitudes at same site of stimulation and recording (p<0.02, n = 5), the variation of peak-amplitude with pulse-width was not significant (n = 4).

## Comparison between two-photon and single-photon for in-depth stimulation

In order to compare the efficacy of two-photon laser beam in stimulating in-depth brain regions in-vivo, with that of single-photon, we first theoretically evaluated light propagation in the brain. Fig. 3a shows results of Monte Carlo simulations of two-photon (870 nm) light propagation in the two-layered cortex for beam (diameter: 60 $\mu$m; laser power: 5 mW) emanating from a fiber having numerical aperture (NA) of 0.15. Propagation of a single-photon (470 nm) fiber-optic beam for the same parameters (beam diameter, NA and laser power) is shown in Fig. 3b. While it is evident that the near-infrared two-photon beam has a higher penetration depth (even at same power levels), this effect is more profound when (damage-threshold) normalized power levels are compared. For in-vivo two-photon stimulation, besides the input beam characteristics (mode profile, beam size, divergence), the forward scattering nature of the medium (brain) further modulates the beam propagation and thus, impact both axial and transverse stimulation-resolution. Fig. S5 shows MC simulation results depicting the effects of different parameters (power, NA, and wavelength) on light propagation in the two-layered cortex. For both FO-TPOS (870 nm) and FO-SPOS (470 nm), higher penetration depth was observed with increasing laser beam power (1 to 50 mW) as shown in the XZ-distribution of power density (W/cm$^2$) in Fig. S5. In our experiments, ~50 times higher average power density was used in the case of FO-TPOS as compared to that of FO-SPOS. Therefore, when results of FO-TPOS with average laser beam power of 50 mW is compared with 1 mW of FO-SPOS, ~3 times higher penetration depth was observed in case of FO-TPOS for same fiber delivery geometry.

Next, in order to experimentally evaluate the in-depth activation efficacy of FO-TPOS in in-vivo transgenic mouse models, we employed the set-up shown in Fig. 2a, with the electrode tip at varying axial distances from the stimulating fiber-tip. The fiber tip was placed in cortical layer-1 of the transgenic mice with the electrode being translated stereotaxically to different depths of the brain. To evaluate the ChR2-YFP expression in the targeted deep-brain regions of Thy1-ChR2-YFP transgenic mice, confocal immunofluorescence imaging of these regions was conducted. Fig. 3c shows expression of reporter-molecule (YFP) at a depth of ~3 mm in the ventral tegmental area (VTA). Co-immunostaining for tyrosine hydroxylase (TH) and the composite images are shown in Fig. 3d & e respectively. The colocalization of tyrosine hydroxylase (TH) and ChR2-YFP in neurons are marked by arrows. These results demonstrate that greater than 90% of YFP positive neurons in the VTA of Thy1-ChR2-YFP transgenic mice were also positive for tyrosine hydroxylase (Fig. 3e). In Fig. 3f, we show fiber-optically stimulated raw spiking activity from excitatory neurons at three different axial distances (1, 2 and 3 mm) from the fiber-tip carrying the two-photon stimulation beam (870 nm). Though with increasing depths, the peak amplitude decreased, spiking signal could still be detected at large depths (3 mm).

In contrast to two-photon activation, no spiking signal was observed at large depths (3 mm) when single-photon beam was used for stimulating the same sites as used for two-photon. Comparison of single and two-photon activated depth-dependent spike-amplitude is shown in Fig. 3 g. In Fig. 3 h, we show the significant variation of peak amplitude of two-photon (850 nm) activated local electrical activity as a function of depth at different power densities (p<0.001, n = 5). Similar depth-dependent variation of peak amplitude was observed for 870 nm (Fig. S6a) and 900 nm (Fig. S6b) at different power densities. Fig. 3i shows comparison of peak-amplitude at different depths due to *in-vivo* FO-TPOS for two different wavelengths (850 and 870 nm) at same average power density (0.024 mW/$\mu m^2$). For all the power densities, while there is no significant decrease in peak amplitude from 1 to 2 mm, significant (p<0.001, n = 5) decrease in signal amplitude was observed when moving from 2 to 3 mm depth. This is true for all the near-infrared wavelengths reported here. Fig. S6c & d respectively shows the histogram of peak-amplitude as a function of power densities of *in-vivo* FO-TPOS at different depths for 850 nm and 870 nm.

## Fiber-optic two-photon optogenetic stimulation is direct stimulation

To verify whether fiber-optic two-photon beam can directly stimulate neurons *in-vivo* at large depth, analysis of FO-TPOS evoked spike latency (from the onset of light) at different depths from fiber-tip was carried out. The experimental setup in Fig. 2a was used to compare the latency values for two-photon with that of single-photon. Figs. 4a and b show schematics of in-depth stimulation limit of fiber-optic single-photon and two-photon beam respectively. In single and two-photon stimulation modalities, both direct and indirect (post-synaptic) stimulations of neurons at different cortical layers (marked by blue line in Fig. 4 a & b) is possible. For single-photon stimulation, since the blue light is attenuated within 1 mm of propagation in cortical tissue (Fig. S5), it is suspected that the spikes from larger depths are mostly due to indirect stimulation. However, at depth of 3 mm from the fiber-tip, no detectable signal was observed (Fig. 4c) in case of blue-light stimulation (473 nm, 4 mW). In contrast to this, two-photon near-infrared beam has penetration depth of ~3 mm and therefore can directly elicit action potentials in neurons at larger depths (2–3 mm) along with the indirect post-synaptic spikes. Raw electrical recording at 3 mm depth for two-photon (870 nm, 80 mW) stimulation is shown in Fig. 4d. In order to estimate the latency of the spikes evoked by FO-SPOS and FO-TPOS at different depths, very short (1 ms) stimulation pulses were used for these experiments and the raw spikes were overlaid with the shutter-synchronization pulses.

Fig. 4e shows the overlay of raw spikes (red profiles) with shutter (FO-TPOS ON) opening (black profiles) at different depths for two different wavelengths (850 and 870 nm). The representative single photon signals (blue profiles) at different depths are also shown. The comparison of values for spike latencies at different depths stimulated by single photon (473 nm) and two-photon (870 nm) fiber-optic beams is shown in Fig. 4f. Though the mean value of spike latency increased with increasing depth for FO-TPOS, it was not statistically significant and therefore should have primarily originated from direct stimulation. The slight increase can be attributed to the contribution from spikes generated by indirect stimulation. However, the statistically significant (p<0.001, n = 5) increase in latency values of spikes from 2 mm depth as compared to 1 mm depth in case of FO-SPOS (Fig. 4f) indicates that most of the spikes are generated by post-synaptic cortical circuitry rather than direct stimulation by the blue laser beam. At higher NIR wavelengths, direct photothermal stimulation of neurons may occur due to absorption by water. To check if latency is wavelength-dependent (owing to the contribution from photothermal stimulation, if any), we carried out comparative studies using three different FO-TPOS wavelengths (850, 870 and 900 nm). Fig. 4 g shows the histogram of latency of spikes at different two-photon wavelengths for three depths (1, 2 and 3 mm). The latency at 900 nm was slightly higher that that due to other two wavelengths, but statistically insignificant. We further studied the effect of average power density on the latency of two-photon (870 nm) evoked spikes. The variation of latency values as a function of depth for three different two-photon average power densities is shown in Fig. 4 h. All the mean latency values are found to be below 6 ms, implying major contribution from direct stimulation. The increase in latency values at 3 mm depth was found to be statistically significant and may be attributed to some contribution from post-synaptic spiking.

## Immunohistochemistry confirms in-depth fiber-optic two-photon optogenetic stimulation

To further confirm whether FO-TPOS was modulating the activity of dopaminergic cells in VTA, we assessed the expression of c-Fos in the VTA using immunohistochemistry (IHC). One hour after fiber-optic two-photon stimulation, the mice were sacrificed and perfused intracardially. The brains were extracted and processed for IHC for c-Fos using procedures described in literature [38,39]. Equally-spaced sections spanning the entire region of interest were analyzed. We used separate groups of non-ChR2 expressing mice to run through the stimulation procedure as a control comparison group for the IHC. Fig. 5a shows confocal YFP-immunostained images of VTA of Thy1-ChR2-YFP transgenic mice. Co-immunostaining for tyrosine hydroxylase (TH, Fig. 5b) showed co-localization (marked by arrows) of YFP in dopaminergic cells (Fig. 5c). Fig. 5d and e shows zoomed immunostained images for TH and c-Fos respectively. The increased activity of the dopaminergic neurons of VTA due to fiber-optic two-photon optogenetic stimulation (FO-TPOS) is evidenced by an increase in c-Fos expression (Fig. 5e) in these cells (Fig. 5f). In Fig. 5 d to f, few cells have been marked by the envelope of the FO-TPOS beam profile (as predicted by MC simulation, Fig. S5).

## Discussion

The two-photon activation method reported here enabled *in-vivo* optogenetic stimulation at depth of 3 mm in the brain. Thus, the fiber delivering the FO-TPOS beam could be positioned in superficial cortical layer, leading to minimal invasion into the brain. Though from the in-vitro studies, we ruled out the contribution of direct (photothermal or photomechanical) stimulation, we are intrigued by the ability of the NIR ultrafast multimode beam in stimulating deep brain regions. We hypothesize that two possibilities can enhance the stimulation efficacy of the NIR fiber-optic beam. First, the ChR2 may have non-zero (but small) absorption in NIR spectrum. Though single-photon ChR2 activation spectrum has been characterized in visible range, further studies are required to measure single-photon NIR activation of ChR2. This would certainly require intense cw light source (average power higher than used in studies reported here) owing to the low activation efficiency in the NIR. Secondly, other molecules such as the cofactor (ATR) can absorb NIR light and transfer energy to the opsin by the process of resonance energy transfer (RET).

This non-linear optogenetic approach, combined with electrophysiology and behavioral readout(s) will provide a unique opportunity to dissect the functional neuronal circuitry of deep brain regions. The efficacy of the two-photon beam can be further improved by controlling the divergence of the beam emanating from the fiber. Since the two-photon process is non-linear in nature, focused near infrared activation can lead to highly localized activation of the specific region of interest, which may be located deep in the ventral portion of the midbrain. Our previous studies (both experimental and Monte Carlo simulation [40,41,42]) show that a non-diffracting optogenetic Bessel beam is more effective for in-depth stimulation than a classical (Gaussian) beam [43]. Fig. S7 shows the comparison of microscopic two-photon setup with defocused and focused fiber-optic two-photon optogenetic stimulation. For *in-vivo* two-photon activation, use of an axicon tipped fiber [44] will allow generation of Bessel beam for even better penetration. The propagation distance of the stimulating Bessel beam and the stimulation volume can be controlled by the cone angle of the fiber tip. The use the non-diffracting ultrafast Bessel beam for FO-TPOS of excitable cells will allow minimally invasive stimulation with improved spatial resolution. This can significantly enhance previous work using single-photon technology for mapping neural circuitry. Further, this method can be easily integrated with the fiber-optic non-linear endoscopy techniques [45] to allow both two-photon stimulation and optical imaging of neural activity *in-vivo*.

## Conclusions

Though we characterized two-photon activation efficacy at different near-infrared laser parameters, further studies are required to optimize the fiber-optic two-photon stimulation strategy. For example, the large cross-section of ChR2 should allow use of nanosecond or even microsecond compact near-infrared (NIR) sources for FO-TPOS. Further, by selection of right wavelength of the two-photon light source for other opsins such as NpHR (chloride channel), the *in-vivo* FO-TPOS method can be useful for inhibiting neural activity. In addition to the demonstrated in-depth stimulation capability of FO-TPOS, high spatial precision in two-photon optogenetic activation should be possible *in-vivo*, by virtue of non-linear nature of ultrafast light-matter interaction. Modulation of deep brain regions via FO-TPOS as demonstrated by us, will lead to better understanding of neural circuitry because the technology permits noninvasive and more precise anatomical delivery of stimulation.

## Supporting Information

**Figure S1  Fiber-optic two-photon optogenetic stimulation set up and characterization.** (a) Patch clamp set up for FO-TPOS. FSL: Tunable Ti: Sapphire Laser; BE: Beam Expander; S: Shutter; NDF: Neutral density filter; L: Lens for fiber coupling; FL: Fluorescence excitation source; Ex: Excitation Filter; Em: Emission Filter; MO: Microscope Objective; CL: Condenser lens; DM: Dichroic Mirror; M: Mirror; HL: Halogen Lamp. (b) Variation of the width of the ultrafast pulse from the multimode fiber as a function of laser power. (c) *In-vivo* electrophysiology set up for FO-TPOS. (d) Typical transverse beam profile emanating from the multimode fiber; (e) Time-lapse (1 sec) intensity profiles (in different colors) along a line drawn across the beam profile shown in d.
(DOCX)

**Figure S2  Estimation of two photon absorption cross section for ChR2.** (a–b) Different traces of photocurrent in ChR2-sensitized HEK cells induced by fiber-optic near-infrared stimulation using 250 fs pulsed laser beam (870 nm, 80 MHz) at average intensity of $0.02 \text{ mW}/\mu\text{m}^2$. The fitted data (using Eq. 1) is overlaid (red traces) over the measured photocurrent. The fitted parameters $\tau_1$ and $\tau_2$ are used to calculate the two-photon absorption cross section.
(DOCX)

**Figure S3  Wavelength-dependent fiber-optic two-photon optogenetic stimulation *in-vivo*.** (a) Raw spiking activity during *in-vivo* electrophysiology subsequent to fiber-optic two-photon optogenetic stimulation (FO-TPOS) at different wavelengths (average incident laser power: 80 mW, 5 Hz), (b) Peak voltage vs. wavelength of *in-vivo* FO-TPOS at two different laser power densities, (c) Firing rate (spikes per second) vs. wavelength of *in-vivo* FO-TPOS.
(DOCX)

**Figure S4  Fiber-optic two-photon optogenetic stimulation of negative control.** Direct two-photon illumination of micropipette-electrode separated by 1 mm (a) no media and (b) in phosphate buffer saline. (c) Recording from micropipette-electrode during two-photon illumination in dead brain.
(DOCX)

**Figure S5  Monte Carlo simulation of light propagation in two-layered cortex for comparison of fiber-optic two-photon optogenetic stimulation vs fiber-optic single-photon optogenetic stimulation.** Effects of numerical aperture (NA) of optical fiber (0.15, 0.22, and 0.3) and laser beam power (1, 5, and 50 mW) on XZ-distribution of power density (W/$\text{cm}^2$) shown for near-infrared (870 nm) as well as blue (470 nm). Parameters for the MC simulation are listed in Table S1.
(DOCX)

**Figure S6  Intensity and depth-dependent fiber-optic two-photon optogenetic stimulation.** Variation of peak amplitude of two-photon activated electrical recording as a function of depth at different power densities (a) 870 nm, (b) 900 nm. Histogram of peak-amplitude as a function of power densities of *in-vivo* fiber-optic TPOS at different depths. (c) 850 nm, (d) 870 nm.
(DOCX)

**Figure S7  Comparison of microscopic two-photon setup with defocused and focused fiber-optic two-photon optogenetic stimulation.**
(DOCX)

**Table S1  Parameters for Monte Carlo simulation.** Beam power is 5 mW, diameter is 60 µm and NA is 0.15. Where, $\mu_a$ = Absorption coefficient; $\mu_s$ = Scattering coefficient; g = anisotropy factor; n = refractive index, and d = thickness of cortical layers.
(DOCX)

**Movie S1  Fiber-optic multimode two-photon excitation beam profile over a period of time.**
(AVI)

**Movie S2  Green fluorescence emitted from polystyrene particles upon excitation by a moving multimode fiber-optic beam.**
(AVI)

## Acknowledgments

The authors would like to thank R. Patil, and A. Kanneganti for assistance during preliminary experiments.

## Author Contributions

Analyzed the data: KD LG SKM LP. Contributed to the writing of the manuscript: KD LG SKM LP. Conceived and supervised the project:

SKM. Performed the Monte Carlo simulations: TL. Performed patch clamp experiments: LG KD. Performed the in-vivo two-photon experiments: LG SS KD SKM TSD LP. Performed the histological sectioning, immunohistochemistry and confocal microscopy experiments: SAMB LG LP.